\documentclass[11pt]{article}
\pdfoutput=1
\usepackage{soul}
\usepackage{jheppub}
\usepackage{amsfonts}
\usepackage{amsmath}
\allowdisplaybreaks[4]         
\usepackage{amssymb}   
\usepackage{euscript}       
\usepackage{xcolor}          
\usepackage{tensor}     
\usepackage{caption}
\usepackage{subcaption}   
\usepackage{graphicx}
\usepackage{float}
\usepackage{braket}
\usepackage{appendix}
\usepackage{cases}
\usepackage{mathtools}
\usepackage{docmute}
\usepackage{mathrsfs}
\usepackage{subfiles}
\usepackage{amsthm}
\usepackage{fancyhdr}
\usepackage{enumerate}
\usepackage{paralist}
\usepackage{color}

\usepackage[T1]{fontenc}
\usepackage[utf8]{inputenc}

\title{Cosmic Inflation From Regular Black Holes}

\author[a]{Kensuke Sueto}
\author[b]{, Riku Yoshimoto}
\author[c]{, Pablo A. Cano}

\affiliation[a]{Department of Physics, Graduate School of Science, Osaka Metropolitan University, Osaka 558-8585, Japan}
\affiliation[b]{Department of Physics, Graduate School of Science, Nagoya University, Nagoya 464-8602, Japan}
\affiliation[c]{Departamento de Física, Universidad de Murcia, Campus de Espinardo, 30100 Murcia, Spain}

\emailAdd{ksueto@omu.ac.jp}
\emailAdd{yoshimoto.riku.d1@s.mail.nagoya-u.ac.jp}
\emailAdd{pablocano@um.es}

\date{\today}

\abstract{We study braneworld cosmology in quasi-topological gravity (QTG) with an infinite tower of higher-curvature terms, focusing on the case in which the bulk admits regular black hole solutions. We derive the $\mathbb{Z}_2$-symmetric junction conditions for a FLRW brane moving in a static, spherically symmetric bulk geometry, and obtain the corresponding modified Friedmann equations for the scale factor.  
We prove that, in the small scale factor regime, the brane generically approaches a de Sitter phase characterized solely by the length scale $\sqrt{\alpha}$ of the higher-derivative terms, while the standard Einstein-gravity braneworld dynamics is recovered in the low-energy regime. We further provide a universal estimate for the number of e-folds of the de Sitter phase in terms of the ratio between the black hole scale and the scale of new physics $r_g/\sqrt{\alpha}$. The inflationary regime is fully independent of the brane matter content and hence avoids the problem of trans-Planckian matter densities. Numerical integrations for explicit regular bulk solutions (Dymnikova-like and Hayward black holes) confirm these estimates and illustrate how the bulk black hole sector controls the onset and termination of inflation.
This framework leverages the powerful properties of QTGs, defined only in $D\ge 5$, to study consequences for a four-dimensional universe. 
}

\begin{document}
\maketitle

\section{Introduction}

General relativity (GR) has successfully passed a wide range of experimental and observational tests and is therefore widely regarded as the correct low-energy description of gravity. Nevertheless, there are important reasons to regard GR as an effective theory of gravity. The long-standing problem of reconciling GR with quantum field theory is perhaps the most clear indication that GR is not the complete story, but there are at least two additional reasons. First, under standard energy conditions for the matter sector, spacetime singularities arise generically in gravitational collapse, leading to a breakdown of the theory \cite{Penrose:1964wq,Hawking:1970zqf}. Second, there are open problems in cosmology that could be addressed by a modification of Einstein's theory \cite{Clifton:2011jh,Capozziello:2011et}. One of those open questions is cosmic inflation \cite{Guth:1980zm,Linde:1981mu,Albrecht:1982wi,Linde:1983gd}, which is invoked to account for the observed large-scale homogeneity and near-flatness of the universe. 

The standard inflationary paradigm cannot be explained by GR alone, and it requires additional model building \cite{Bassett:2005xm,Linde:2007fr}. However, the precise mechanism that drove the early expansion of the universe is still unknown. In many models, inflation is driven by a scalar inflaton field. Achieving a sufficiently long inflationary phase and a graceful exit often requires tuning of the inflaton potential, and no inflaton has been directly observed. These issues motivate the exploration of alternative scenarios. Since the very early universe is a period of extremely high curvature, ultraviolet (UV) modifications of gravity, such as higher-curvature effective actions, are likely to play a role.
In fact, one of the most successful models of inflation, Starobinky's model \cite{Starobinsky:1980te}, is based on the addition of quadratic curvature terms (in particular, an $R^2$ term). However, this model is still secretly driven by a scalar inflaton field. Other inflationary models with additional higher-curvature corrections have also been proposed in the literature, \textit{e.g.} \cite{Berkin:1990nu,Biswas:2012bp,Bonanno:2015fga,Koshelev:2016xqb,RomeroCastellanos:2018inv,Carloni:2018yoz,Cano:2020oaa,Bianchi:2025tyl}.

In this paper, we are interested in higher-curvature modifications of GR that can be understood as effective field theories, and that therefore do not introduce new dynamical degrees of freedom. 
In this context, an attractive class of higher-curvature theories is provided by the generalized quasi-topological gravities (GQTGs) \cite{Bueno:2016xff,Hennigar:2017ego,Bueno:2017sui,Ahmed:2017jod,Bueno:2019ycr,Bueno:2022res}. These satisfy a number of properties that make them appealing for different applications.  These theories admit non-hairy static, spherically symmetric black holes characterized by a single metric function satisfying a second-order differential equation in the Schwarzschild gauge. Moreover, up to perturbative field redefinitions, any effective action built solely from contractions of the metric and the Riemann tensor can be expressed in a GQTG basis \cite{Bueno:2019ltp}.  In addition, a subset of GQTGs gives rise to second order Friedmann equations for FLRW metrics, allowing one to study their impact in cosmology in a non-perturbative way \cite{Arciniega:2018fxj,Arciniega:2018tnn,Cisterna:2018tgx,Moreno:2023arp}.

Motivated by these structural properties, GQTGs have been explored in early-universe cosmology \cite{Arciniega:2018fxj,Arciniega:2018tnn,Cisterna:2018tgx}. Indeed, when an infinite tower of higher-curvature terms is included, an early-time de Sitter phase can replace the big bang singularity. In these models, inflation can arise without introducing an inflaton field; instead, it is driven by higher-curvature corrections. This mechanism is referred to as geometric inflation. Furthermore, at sufficiently large scale factor the higher-curvature contributions become subdominant, allowing the evolution to approach the Einstein-gravity regime and facilitating an exit from the quasi-de Sitter phase. These features suggest that geometric inflation may offer a unified framework for singularity avoidance and an early quasi-de Sitter phase, followed by a transition toward standard cosmological evolution. However, these scenarios suffer from a key drawback: achieving the observationally required number of e-folds typically requires trans-Planckian matter densities \cite{Edelstein:2020nhg}. This limitation leaves the phenomenology unsettled and motivates alternative constructions in which early-time acceleration can be achieved without invoking trans-Planckian matter. In addition, although GQTGs can be formulated in $D=4$, explicit black hole solutions with an infinite tower of these corrections have not yet been constructed. Therefore, GQTGs leave at least two open questions: (i) whether a realistic early-universe cosmology can be implemented in four dimensions without invoking trans-Planckian matter, and (ii) whether black hole singularities are resolved within an infinite-tower setup.

Two particularly relevant subclasses of GQTG include quasi-topological gravity (QTG) and Birkhoff quasi-topological gravity (Birkhoff-QTG)\footnote{We refer to \cite{Bueno:2025qjk} for a clear explanation of the different notions of quasi-topological gravity.}. In this article, we focus on the Birkhoff-QTG. This family of theories is defined so that the field equations remain second order in general spherically symmetric spacetimes, and they satisfy a Birkhoff theorem: all vacuum spherically symmetric solutions are static and uniquely characterized by a single continuous integration constant, together with a possible additional discrete parameter \cite{Bueno:2024dgm,Bueno:2025qjk}. Remarkably, when an infinite tower of higher-curvature corrections is included, singularities are removed and the vacuum solutions of these theories automatically become regular black holes under very broad conditions \cite{Bueno:2024dgm}. Furthermore, these regular black holes can dynamically form from gravitational collapse in close analogy to standard collapse scenarios in Einstein gravity \cite{Bueno:2024eig,Bueno:2024zsx, Bueno:2025gjg}. In this respect, this theory provides a concrete and technically tractable mechanism for singularity resolution in a purely gravitational setting, and many follow-up studies have subsequently been proposed \cite{Konoplya:2024kih, Aguayo:2025xfi, Fernandes:2025fnz,Fernandes:2025eoc, Hennigar:2025ftm, Arbelaez:2026eaz}.

A well-known issue is that (polynomial\footnote{One can find four-dimensional theories with similar properties to QTG if one allows for Lagrangian densities which are non-polynomial in curvature invariants \cite{Colleaux:2017ibe,Bueno:2025zaj,Borissova:2026wmn,Borissova:2026krh,Borissova:2026klg}. However, this type of Lagrangians is not so well motivated from the point of view of EFT.}) QTG becomes trivial in four dimensions, reducing to GR. Since our universe is effectively four-dimensional, this prevents a direct phenomenological application of Birkhoff-QTG as fundamental four-dimensional theories. A minimal way to reconcile this with four-dimensional cosmology is to adopt a braneworld perspective \cite{Randall:1999ee, Randall:1999vf,Binetruy:1999ut, Ida:1999ui, Shiromizu:1999wj}: the bulk is five-dimensional and governed by a Birkhoff-QTG, while the brane is a four-dimensional hypersurface moving in the bulk spacetime. The braneworld construction is well motivated in string/M-theory \cite{Horava:1995qa,Randall:1999ee, Randall:1999vf} as a mechanism to explain the four dimensions of our universe, but it can generally be applied to any higher-dimensional theory of gravity. In brane cosmology with an (A)dS–Schwarzschild type bulk, the bulk black hole mass enters in the effective Friedmann equation on the brane as a dark radiation term, whose energy density is proportional to the bulk mass parameter \cite{Ida:1999ui}. This establishes a direct channel through which the bulk black hole sector can control early-time brane dynamics.

In this work, we develop and analyze a braneworld cosmology in which a four-dimensional FLRW brane evolves in a five-dimensional bulk governed by Birkhoff-QTG admitting regular black hole solutions. We derive the resulting effective cosmological evolution on the brane and show that the regularity of the bulk black hole sector leads to a de Sitter phase characterized by the length parameter of Birkhoff-QTG, $\alpha$. We show that this setup avoids the trans-Planckian matter requirement encountered in previously studied infinite-tower GQTG cosmologies. This article is organized as follows. In Section \ref{sec-review}, we review Birkhoff-QTG, focusing on its definition and its regular black hole solutions. In Section \ref{sec-braneworld} we derive $\mathbb{Z}_2$-symmetric junction conditions for a brane moving on a regular black hole solution and obtain the resulting modified Friedmann equations for the scale factor. We analyze two specific Birkhoff-QTG theories (admitting Dymnikova-like and Hayward regular black holes), prove that the de Sitter regime arises universally, and present an analytic estimate for the number of e-folds. We also include numerical solutions of the modified Friedmann equations. Section~\ref{sec-Conclusion} contains our conclusions and outlook.

\section{Bulk theory: Birkhoff-QTG and regular black hole solutions}
\label{sec-review}
In this section, we briefly review Birkhoff quasi-topological gravity (Birkhoff-QTG) and its regular black hole solutions, and we summarize the junction conditions required for the braneworld setup.
\subsection{Definition of Birkhoff-QTG}
Birkhoff-QTG is a higher-curvature gravity theory whose Lagrangian is constructed solely from the metric and the Riemann tensor, and which is defined by the property that, for a general spherically symmetric spacetime, the field equations reduce to at most second-order differential equations.  A crucial observation is that, while several Lagrangian densities of this type may exist at each curvature order, they all give rise to the same equations of motion on spherical symmetry; therefore, for the purposes of studying spherically symmetric solutions, it is enough to consider one representative Birkhoff-QTG density at each order $n$, that we denote by  $\mathcal{Z}_{(n)}$. The explicit expressions of these densities are lengthy,  but they can be found in \cite{Bueno:2024zsx, Bueno:2025gjg}, where it is furthermore proven that these densities exist at all curvature orders in $D\ge 5$.

Throughout this article, we restrict our attention to $D\geq 5$ bulk dimensions and we consider the theory 
\begin{align}
  \label{eq-Quasi-topological-action}
    I=\frac{1}{16\pi G_N}\int{d^Dx\sqrt{-g}\left[R-\frac{\Lambda}{(D-1)(D-2)}+\sum_{n=2}^{n_{\max}}\alpha_n\mathcal{Z}_{(n)}\right]}\, ,
\end{align}
where we are including the Einstein-Hilbert term, a cosmological constant $\Lambda$ and higher-order Birkhoff-QTG densities up to a maximum order $n_{\rm max}$. For the purposes of braneworld cosmology, we are interested in the case of $\Lambda<0$, and we will also take the limit $n_{\rm max}\to \infty$. While we keep $D$ general throughout most of the article, we are most interested in the case of $D=5$ bulk dimensions, giving rise to a four-dimensional braneworld. 

We focus on analyzing the dynamics of Eq.~\eqref{eq-Quasi-topological-action} on spherical symmetry. 
Taking the general spherically symmetric spacetime metric as
\begin{align}
  \label{eq-general-spherical-metric}
  ds^2=\gamma_{AB}dx^Adx^B+\varphi^2(x)d\Omega_{D-2}^2\quad (A,B=1,2),
\end{align}
  where $\gamma_{AB}$ denotes a two-dimensional metric and $d\Omega_{D-2}^2$ denotes the metric of the $(D-2)$-sphere, the action  Eq.~\eqref{eq-Quasi-topological-action} reduces to an effective two-dimensional action for $\gamma_{AB}$ and the scalar field $\varphi$, 
\begin{align}
  I&=\frac{1}{16\pi G_N}\int{d^Dx \sqrt{-g}\mathcal{L}(g^{ab},R_{abcd})}=\frac{(D-2)\Omega_{D-2}}{16\pi G_N}\int{d^2x\sqrt{-\gamma}\mathcal{L}_{2d}(\gamma_{AB},\varphi)}\, .
  \end{align}
 The two-dimensional Lagrangian belongs to the Horndeski class \cite{Horndeski:1974wa} --- signaling that the theory possesses second-order equations of motion --- and it reads explicitly 
\begin{align}
  \label{eq-QTG-2D-Lagangian}
  \mathcal{L}_{2d}&=G_2(\varphi,X)-\square\varphi G_3(\varphi,X)+G_4(\varphi,X)R-2G_{4,X}(\varphi,X)\left[(\square \varphi)^2-\nabla_A\nabla_B\varphi\nabla^A\nabla^B\varphi\right]\, , 
    \end{align}
 where    
\begin{align}
         X\coloneqq \nabla_A\varphi \nabla^A\varphi\, ,\qquad
          \psi\coloneqq \frac{1-X}{\varphi^2}\, ,\qquad
          G_{4,X}\coloneqq \partial_XG_4\, ,
\end{align}
and $G_{2,3,4}$ are the functions   
\begin{align}
  \begin{cases}
    G_2(\varphi,X)&= \varphi^{D-2}((D-1)h(\psi)-2\psi h'(\psi))\, ,\\
    G_3(\varphi,X)&= 2\varphi^{D-3}h'(\psi)\, ,\\
    G_4(\varphi,X)&= -\frac{\varphi^{D-2}}{2}\psi^{\frac{D-2}{2}}\int{d\psi \frac{h'(\psi)}{\psi^{\frac{D}{2}}}}\, .
  \end{cases}
\end{align}
A notable feature is that these functions are not all arbitrary, as they are fixed in terms of the \emph{characteristic function} $h(\psi)$ of the corresponding QTG theory, which is given by\footnote{Here $\alpha_{1}=1$ corresponds to the contribution of the Einstein-Hilbert term.}
\begin{equation}
          h(\psi)\coloneqq -\Lambda+\sum_{n=1}^{n_{\max}}{\alpha_n\psi^n}\, .
\end{equation}
We remark that, the only difference with respect to the derivation in \cite{Bueno:2024zsx,Bueno:2024eig} is that the cosmological constant enters as a constant term in $h(\psi)$. We note that for finite $n_{\rm max}$ the characteristic function is a polynomial, but the resummation of the infinite series for $n_{\rm max}\to\infty$ yields other types of functions with qualitatively different features. This is exploited for the construction of regular black hole solutions. 

Interestingly, the 2-dimensional Lagrangian Eq.\eqref{eq-QTG-2D-Lagangian} is identical to the one obtained from Lovelock gravity \cite{Kunstatter:2015vxa}. The main difference though is that in the case of QTG, one can take $n_{\rm max}$ to be arbitrarily large, while in Lovelock gravity one has $n_{\rm max}\le \lfloor D/2\rfloor$, which does not allow for singularity resolution. 
Ref.~\cite{Kunstatter:2015vxa} observed that, if one could take $n_{\rm max}\to \infty$ by introducing dimensionally-reduced ``regularized'' Lovelock densities, then the corresponding theory would allow for regular black hole solutions. However, it turns out this regularization procedure only works for planar black holes \cite{Fernandes:2025fnz}. Quasitopological gravities overcome this problem as they exist at any order in any given dimension $D\ge 5$. 

In passing, let us note that more general classes of theories admitting regular black hole solutions can be achieved if one allows the Lagrangian to be a non-polynomial function of curvature invariants   \cite{Colleaux:2017ibe,Bueno:2025zaj,Borissova:2026wmn,Borissova:2026krh,Borissova:2026klg}. 
    
\subsection{Black hole solutions}
To find black hole solutions, we take the two-dimensional metric as 
    \begin{align}
      \label{eq-two-dim-metric}
      \gamma_{AB}dx^Adx^B=-N^2(r,t)f(r,t)dt^2+\frac{dr^2}{f(r,t)},
    \end{align}
and $\varphi(x)=r$. In the absence of matter, the equations of motion for the metric functions $f(r,t)$ and $N(r,t)$ reduce to \cite{Bueno:2024zsx, Bueno:2024eig}
      \begin{align}
        f&=f(r)\, ,\\
        N&=N(t)\, ,\\
        h(\psi)&=\frac{2M}{r^{D-1}}\, ,\quad \text{where}\quad \psi=\frac{1-f}{r^2} \label{eq-h-Regular-black-hole},
      \end{align}
and where $M$ is an integration constant proportional to the mass. 
Choosing a new time coordinate as $N(t)dt\rightarrow dt$, this spacetime becomes static, expressed as 
\begin{align}
  \label{eq-ff-1}
  ds^2=-f(r)dt^2+\frac{dr^2}{f(r)}+r^2d\Omega_{D-2}^2\, ,
\end{align}
where the metric function $f(r)$ is determined by the algebraic equation Eq.~\eqref{eq-h-Regular-black-hole}.
Thus, Birkhoff's theorem holds, and the Schwarzschild-(A)dS solution is recovered when $\alpha_n=0$ for all $n\geq 2$. 

In order to examine the existence of curvature singularities, we consider the behavior of $f(r)$ near $r=0$. If the series is truncated at some $n=n_{\rm max}$, in the vicinity of $r=0$, $f(r)$ behaves as 
\begin{align}
  f(r)=1- \left(\frac{2M}{\alpha_{n_{\rm max}}}\right)^{\frac{1}{n_{\rm max}}}r^{2-\frac{D-1}{n_{\rm max}}}\, .
\end{align}
Therefore, to obtain regular solutions, one must take the limit $n_{\max}\rightarrow\infty$. In terms of $h(\psi)$, it is sufficient for this function to be a one-to-one function between a compact interval containing $0$ and the full real line, $h: [\psi_{1},\psi_{0}]\to \mathbb{R}$, with $\psi_{1}<0<\psi_{0}$. 
In terms of the coupling constants, this can be achieved by imposing \cite{Bueno:2024zsx}
\begin{align}
  \label{condition-to-parameter}
\alpha_n\geq0\ (\forall n)\,,\quad \lim_{n\rightarrow\infty} {\alpha_n}^{\frac{1}{n}}=\alpha>0\, ,\quad \text{and} \quad \sum_{n=1}^{\infty}\alpha_n=\infty\, . 
\end{align}
Again, these are sufficient but not necessary conditions in order to ensure regularity of the solution --- there are more choices that lead to singularity-free black holes. 
The conditions Eq.~\eqref{condition-to-parameter} imply that $h(\psi)$ diverges at $\psi_0=1/\alpha$. This means that $\psi\to 1/\alpha$ when $r\to 0$, and leads to the universal behavior
\begin{align}
  \label{eq-small-r-f}
  f(r)= 1-\frac{r^2}{\alpha}+\ldots
\end{align}
in the vicinity of $r=0$. Therefore, if Eq.~\eqref{condition-to-parameter} is satisfied, the black hole singularity is replaced by a de Sitter core, independently of the value of the mass parameter $M$. We adopt the conditions Eq.~\eqref{condition-to-parameter} throughout this article.

For later use, we present two examples of regular black holes corresponding to specific choices of the coupling constants. 

  \subsubsection*{Dymnikova-like black hole}
  For the choice $\alpha_n=\frac{\alpha^{n-1}}{n}$, one finds $h(\psi)=-\Lambda-\frac{\ln{(1-\alpha\psi)}}{\alpha}$, and the metric function $f(r)$, obtained from solving Eq.~\eqref{eq-h-Regular-black-hole}, reads
  \begin{align}
    \label{eq-Dymnikova-like-f}
    f(r)=1-\frac{r^2}{\alpha}\left(1-\exp{\left[-\alpha\left(\frac{2M}{r^{D-1}}+\Lambda\right)\right]}\right).
  \end{align}
  This metric is reminiscent of the model introduced in \cite{Dymnikova:1992ux} and hence we denote it the Dymnikova-like black hole. We note that this solution has the peculiarity of being $\mathcal{C}^{\infty}$ but non-analytic at $r=0$. 
  
  \subsubsection*{Hayward black hole} 
  For the choice $\alpha_n=\alpha^{n-1}$, one finds $h(\psi)=-\Lambda+\frac{\psi}{1-\alpha\psi}$, and the metric function $f(r)$ is given by
 \begin{align}
  \label{eq-Hayward-f}
  f(r)=1-\frac{2Mr^2+\Lambda r^{D+1}}{(1+\alpha \Lambda)r^{D-1}+2M\alpha }.
\end{align}
This is a $D$-dimensional, (A)dS generalization of the Hayward regular black hole model \cite{Hayward:2005gi}, with the important difference that now  Eq.~\eqref{eq-Hayward-f} is not a phenomenological model, but the unique spherically symmetric vacuum solution of a gravitational theory. 
The thermodynamic properties of these regular black hole solutions with a cosmological constant have been studied in \cite{Hennigar:2025yqm}.

\section{Braneworld cosmology with regular black holes}
\label{sec-braneworld}

\subsection{Junction conditions and effective Friedmann equation}
Braneworld cosmology describes our universe as a brane moving in a higher-dimensional bulk spacetime. Consider a brane $\Sigma$ that divides the spacetime manifold $\mathcal{M}$ into $\mathcal{M}_+$ and $\mathcal{M}_-$. The matching of the metric and equations of motion across this wall is described by the junction conditions. 
In GR, the dynamics of the domain wall are governed by the Israel junction conditions \cite{Israel:1966rt}. The generalization of the Israel junction conditions to Birkhoff-QTG in spherically symmetric spacetimes has already been obtained in Refs.~\cite{Bueno:2024zsx,Bueno:2024eig}, so we summarize the key results next.\footnote{See also \cite{Senovilla:2026fby} for a discussion of junction condition in higher-derivative theories of gravity.} 

We consider the case in which $\Sigma$ is a spherical shell inside a spherically symmetric spacetime. Since our theories satisfy a Birkhoff theorem, the solution inside  and outside  the shell are simply given by Eq.~\eqref{eq-h-Regular-black-hole} with possibly different mass parameters $M_{-}$ and $M_{+}$, respectively. We denote the corresponding metric functions by $f_{\pm}$, so the metric in each side reads
\begin{align}
  \label{eq-ff-2}
  ds_{\pm}^2=-f_{\pm}(r)dt_{\pm}^2+\frac{dr^2}{f_{\pm}(r)}+r^2d\Omega_{D-2}^2\, ,
\end{align}
As the position of the shell depends on time, we can parametrize it on each side of the spacetime by
\begin{equation}
t_{\pm}=T_{\pm}(\tau)\, ,\quad r=a_{\pm}(\tau)\, .
\end{equation}
Thus, the induced metric on $\Sigma$ from $\mathcal{M}_+$ and from $\mathcal{M}_-$ is given by
\begin{align}
  h_{AB}^{\pm}dx^Adx^B=-\left(f_{\pm}\dot{T}_{\pm}^2-f_{\pm}^{-1}\dot{a}_{\pm}^2\right)d\tau^2+a_{\pm}(\tau)^2d\Omega_{D-2}^2\, .
\end{align}
Now, the first junction condition works exactly in the same way as in GR, and it requires the induced metric to be the same on both sides, $ h_{AB}^{+}= h_{AB}^{-}$. Therefore, we have
\begin{equation}
a_{+}(\tau)=a_{-}(\tau)\equiv a(\tau)\, .
\end{equation}
In addition, we can choose $\tau$ to represent the proper time on the shell, which amounts to imposing that 
\begin{equation}\label{Tdoteq}
f_{\pm}\dot{T}_{\pm}^2-f_{\pm}^{-1}\dot{a}^2=1\, .
\end{equation}
Therefore, the induced metric becomes
\begin{align}
  ds^2_{\Sigma}=h_{AB}dx^Adx^B=-d\tau^2+a^2(\tau)d\Omega_{D-2}^2\, ,
\end{align}
which represents a $(D-1)$-dimensional FLRW cosmological metric with spherical spatial sections. 

The second junction condition can be regarded as the equations of motion of the worldbrane, and they take the form
\begin{equation}
\Pi_{AB}^+-\Pi_{AB}^-=-8\pi G_N S_{AB}\, ,
\end{equation}
where $S_{AB}$ is the surface stress-energy tensor on the brane and $\Pi_{AB}^{\pm}$ is a tensor that depends on the geometry of the brane, evaluated on each side of it. The explicit form of this tensor is theory-dependent. For instance, for Einstein gravity, it is given by $\Pi_{AB}=K_{AB}-h_{AB} K$, where $K_{AB}$ is the extrinsic curvature of $\Sigma$ \cite{Israel:1966rt}. For QT gravities, the general form of this tensor is not yet known, but making use of the two-dimensional Horndeski reduction Eq.~\eqref{eq-QTG-2D-Lagangian} and the fact that the junction conditions for Horndeski theory are known \cite{Padilla:2012ze},  Refs.~\cite{Bueno:2024zsx,Bueno:2024eig} manage to derive the form of $\Pi_{AB}$ in spherical symmetry. Employing our notation, the results of \cite{Bueno:2024zsx,Bueno:2024eig} read

\begin{align}\label{eq-junction-component}
\Pi_{\tau\tau}^{\pm}&=\frac{(D-2)}{a}\int_{0}^{a^{\pm}_{n}}dz h'\left(\frac{1+\dot{a}^2- z^2}{a^2}\right)\, ,\\
\Pi_{ij}^{\pm}&=-g_{ij}\frac{1}{(D-2)a^{D-3}\dot{a}}\frac{d}{d\tau}(a^{D-2}\Pi^{\pm}_{\tau\tau})\, ,
\label{eq-junction-component2}
\end{align}
where $i,j$ denote the angular components. In this expression, we have 
\begin{equation}\label{anpmdef}
a^{\pm}_{n}\coloneqq n_{\pm}^b\partial_b a\, ,
\end{equation}
where $n^{a}_{\pm}$ is the normal vector on each side of the brane, normalized by $g_{ab}n^a_{\pm}n^b_{\pm}=+1$. 

\subsubsection*{$\mathbb{Z}_{2}$ symmetry}
 Our analysis describes so far a general spherical thin shell. Motivated by the orbifold construction of \cite{Horava:1995qa}, we now impose $\mathbb{Z}_{2}$ symmetry across the brane, which is a standard assumption in braneworld models \cite{Randall:1999ee, Randall:1999vf,Binetruy:1999ut, Ida:1999ui, Shiromizu:1999wj}. In order to achieve this, on the one hand, we set the mass parameter to be the same on both sides of the shell $M_{+}=M_{-}\equiv M$, which also implies that
 \begin{equation}
 f_{+}=f_{-}=: f\, .
 \end{equation}
 On the other hand, the characteristic feature of $\mathbb{Z}_{2}$ symmetry is that we choose the normal vectors on each side to be opposite to each other \cite{Maartens:2010ar}, 
 \begin{align}
 n_{-}=-n_{+}=\frac{\dot{a}}{f}\partial_t+f\dot{T}_{-}\partial_r\, ,
\end{align}
and taking into account Eq.~\eqref{Tdoteq} we have $f\dot{T}_{-}=\sqrt{f+\dot{a}^2}$. We can then compute $a_{n}^{\pm}$ defined in Eq.~\eqref{anpmdef} using that $a=r$, and we get\footnote{In principle, the sign of $a_n$ could flip at points where $f+\dot{a}^2=0$ \cite{Bueno:2024zsx}. However, as we will see later, in the parameter range of interest below (with $\rho\neq0$ and $\sigma\neq 0$), $a_n$ never vanishes, and the sign is fixed.}
 \begin{align}\label{anvalue}
  a_{n}:=a^{-}_{n}=-a^{+}_{n}&=\sqrt{f+\dot{a}^2}\, .
\end{align}
From this, it is immediate to check that the tensor $\Pi_{AB}$ satisfies 
\begin{align}
  \Pi^{-}_{AB}=-\Pi_{AB}^{+}\eqqcolon \Pi_{AB},
\end{align}
and the junction conditions become
\begin{align}
    \label{eq-junction-braneworld}
  \Pi_{AB}=4\pi G_N S_{AB}.
\end{align}
These represent the equations of motion on the brane $\Sigma$. Next, we assume that the matter on the brane is a perfect fluid:
\begin{align}
    S_{AB}=\rho u_{A}u_{B}+P(h_{AB}+u_Au_B)-\sigma h_{AB},
\end{align}
where $u_{A}=-\delta_{A\tau}$ is the four-velocity of the fluid, $\rho$ is its density, $P$ is its pressure, and $\sigma$ is the brane tension\footnote{Observe that $\sigma$ plays the role of a cosmological constant, and we could include it as a component of the pressure and density with $P_{\sigma}=-\rho_{\sigma}=-\sigma$.}, assumed to be constant.  Let us then evaluate the equations of Eq.~\eqref{eq-junction-braneworld}. Taking into account that  $S_{\tau\tau}=\rho+\sigma$, the equation $\Pi_{\tau\tau}=4\pi G_N S_{\tau\tau}$ takes the form
 \begin{align}
    \label{eq-braneworld-EOM-1}
    \frac{D-2}{a}\int_0^{a_n}{dz h'\left(\frac{1+\dot{a}^2-z^2}{a^2}\right)}=4\pi G_N(\rho+\sigma).
  \end{align}
To obtain the other component of the junction conditions, taking the trace of Eq.~\eqref{eq-junction-braneworld} leads to
\begin{align}
h^{\tau\tau}\Pi_{\tau\tau}+h^{ij}\Pi_{ij}&=4\pi G_N(h^{\tau\tau}S_{\tau\tau}+h^{ij}S_{ij})\, ,
\end{align}
since $h^{\tau\tau}(\Pi_{\tau\tau}-4\pi G_N S_{\tau\tau})=0$ holds, we obtain
\begin{align}
 \label{eq-junction-S}
 h^{ij}\Pi_{ij}=4\pi G_Nh^{ij}S_{ij}.
\end{align}
  From the expression of $S_{AB}$, we have $h^{ij}S_{ij}=(D-2)(P-\sigma)$, and from Eq.~\eqref{eq-junction-component2}, Eq.~\eqref{eq-junction-S} can be written as
  \begin{align}
    \label{eq-braneworld-EOM-2}
    -\frac{1}{a^{D-3}\dot{a}}\frac{d}{d\tau}\left[a^{D-3}\int_0^{a_n}{dz h'\left(\frac{1+\dot{a}^2-z^2}{a^2}\right)}\right]=4\pi G_N(P-\sigma).
  \end{align}
The equations Eq.~\eqref{eq-braneworld-EOM-1} and Eq.~\eqref{eq-braneworld-EOM-2} are the junction conditions of the braneworld scenario, and correspond, respectively, to the first and second Friedmann equations for the scale factor. From the two equations, we can derive several interesting properties. 
First, by substituting Eq.~\eqref{eq-braneworld-EOM-1} into Eq.~\eqref{eq-braneworld-EOM-2}, the energy-conservation law 
  \begin{align}\label{conservationeq}
      \dot{\rho}+(D-2)(\rho+P)\frac{\dot{a}}{a}=0
  \end{align}
follows. Therefore, we only need to solve Eq.~\eqref{eq-braneworld-EOM-1} and Eq.~\eqref{conservationeq}, and we do not need to consider Eq.~\eqref{eq-braneworld-EOM-2} as it implied by the two former equations. 

Second, the sign of $a_n$ is fixed. From the junction condition Eq.~\eqref{eq-braneworld-EOM-1}, $a_n=0$ can occur only if $\rho+\sigma=0$. Since we consider $\sigma>0$ and $\rho> 0$, $a_n$ never vanishes and its sign is fixed as in Eq.~\eqref{anvalue}.

Third, we can obtain universal inequalities that hold for any bulk theory that admits regular black hole solutions. We recall that, if we impose the conditions Eq.~\eqref{condition-to-parameter} in the coupling constants, then $h(\psi)$, and consequently, $h'(\psi)$,  diverge at $\psi_0=1/\alpha$. This allows us to derive two inequalities from Eq.~\eqref{eq-braneworld-EOM-1}. To this end, let us consider the argument of $h'$ in the integral of Eq.~\eqref{eq-braneworld-EOM-1}, 
\begin{align}
  \psi(z)=\frac{1+\dot{a}^2-z^2}{a^2}.
\end{align}
We observe that the integral will diverge if $\psi(z)=1/\alpha$ for any $z$ in the integration interval. Therefore, we must have $\psi(z)\leq 1/\alpha$, and in particular $\psi(a_n)\leq 1/\alpha$ and $\psi(0)\leq 1/\alpha$, which yields the inequalities 
    \begin{subequations}
      \begin{align}
      \label{eq-bulk-constraint}
        \frac{1-f}{a^2}&\leq\, \frac{1}{\alpha},\\
      \label{eq-braneworld-constraint}
        \frac{1+\dot{a}^2}{a^2}&\leq\, \frac{1}{\alpha}\, ,
      \end{align}
    \end{subequations}
respectively. These constraints play a central role in discussing the early universe, as we show later. 
In order to study the cosmic evolution in more detail, we consider next three concrete examples of the coupling constants in Birkhoff-QTG.

\subsection{Einstein gravity}
Birkhoff-QTG reproduces Einstein gravity when $\alpha_{n}=0$ for all $n\geq2$.
When this condition is satisfied, we have $h(\psi)=-\Lambda+\psi$, $f(r)=1-\frac{2M}{r^{D-3}}-\Lambda r^2$ and Eq.~\eqref{eq-braneworld-EOM-1} can be written as
  \begin{align}\label{braneworld-Einstein}
    \left(\frac{\dot{a}}{a}\right)^2&=-\frac{f}{a^2}+\frac{16\pi^2G_N^2}{(D-2)^2}(\rho+\sigma)^2\\
    &=-\frac{1}{a^2}+\frac{32\pi^2G_N^2\sigma}{(D-2)^2}\rho+\left(\Lambda+\frac{16\pi^2G_N^2}{(D-2)^2}\sigma^2\right)+\frac{16\pi^2G_N^2}{(D-2)^2}\rho^2+\frac{2M}{a^{D-1}}.
  \end{align}
In the case of $D=5$, we get a four-dimensional braneworld cosmology, and defining the four-dimensional Newton and cosmological constants as 
  \begin{align}
    \frac{8\pi G_4}{3}&\coloneqq \frac{32\pi^2 G_N^2\sigma}{9},\\
    \frac{\Lambda_4}{3}&\coloneqq \frac{\Lambda}{3}+\frac{16\pi^2G_N^2}{9}\sigma^2\, ,
  \end{align}
Eq.~\eqref{eq-braneworld-EOM-1} reproduces the Friedmann equations of braneworld cosmology in Einstein gravity \cite{Ida:1999ui}:
  \begin{align}
    \label{eq-braneworld-cosmology}
    \left(\frac{\dot{a}}{a}\right)^2=-\frac{1}{a^2}+\frac{8\pi G_4}{3}\rho+\frac{\Lambda_4}{3}+\frac{4\pi G_4}{3\sigma}\rho^2+\frac{2M}{a^{4}}.
  \end{align}
The last term $\rho_{\mathrm{M}}=\frac{2M}{a^4}$ represents radiation originating from the bulk black hole mass and is called \emph{dark radiation}.

\subsection{Dymnikova-like scenario}
 If we choose the bulk theory to be the one leading to the Dymnikova-like regular black hole Eq.~\eqref{eq-Dymnikova-like-f}, the corresponding modified Friedmann equation Eq.~\eqref{eq-braneworld-EOM-1} becomes
  \begin{align}
    \label{eq-modified-Friedmann-Dymnikova-like}
    \left(\frac{\dot{a}}{a}\right)^2+\frac{1}{a^2}-\frac{1}{\alpha}=-\frac{1}{\alpha}\exp{\left[-\frac{2M\alpha}{a^{D-1}}-\alpha \Lambda\right]}\cos^2{\left[\sqrt{-\left(\frac{\dot{a}}{a}\right)^2-\frac{1}{a^2}+\frac{1}{\alpha}}\cdot \frac{4\pi G_N(\rho+\sigma)\alpha}{D-2}\right]}.
  \end{align}
 The derivation of this equation involves carrying out the integration in Eq.~\eqref{eq-braneworld-EOM-1}, and is presented in Appendix~\ref{Appendix-proof}. 

In order to understand the effect of the higher-derivative terms, it is interesting to analyze Eq.~\eqref{eq-modified-Friedmann-Dymnikova-like}  in the regimes of large and small scale factors. 
When the scale factor takes a sufficiently large value, the higher-curvature effects are expected to be weak. Therefore, Eq.~\eqref{eq-modified-Friedmann-Dymnikova-like} should reproduce the Friedmann equation in Einstein gravity, Eq.~\eqref{braneworld-Einstein}. To show this expectation, we define the low-energy regime by 
    \begin{align}
      \alpha\left[\left(\frac{\dot{a}}{a}\right)^2+\frac{1}{a}^2\right]\ll 1,\quad \frac{2M\alpha}{a^{D-1}}\ll 1, \quad \frac{4\pi G_N\sqrt{\alpha}}{D-2}\rho \ll 1,\quad \frac{4\pi G_N\sqrt{\alpha}}{D-2}\sigma\ll 1, \quad \alpha\Lambda\ll 1.
    \end{align}
Note that the first three conditions are naturally satisfied as the scale factor takes a sufficiently large value; however, the last two conditions impose restrictions on the parameters $\sigma$ and $\Lambda$, requiring their associated length scales to be much larger than the length parameter of the Birkhoff-QTG $\sqrt{\alpha}$. In the low-energy regime, we find the expansions
  \begin{align}
    \exp{\left[-\frac{2M\alpha}{a^{D-1}}-\alpha\Lambda\right]}
    &\cong 1-\frac{2M\alpha}{a^{D-1}}-\alpha\Lambda,\\
    \cos^2{\left[\sqrt{-\left(\frac{\dot{a}}{a}\right)^2-\frac{1}{a^2}+\frac{1}{\alpha}}\cdot \frac{4\pi G_N(\rho+\sigma)\alpha}{D-2}\right]}
    &\cong 1-\frac{1}{\alpha}\left(\frac{4\pi G_N(\rho+\sigma)\alpha}{D-2}\right)^2\, ,
  \end{align}
and therefore, Eq.~\eqref{eq-modified-Friedmann-Dymnikova-like} becomes
  \begin{align}\notag
    \left(\frac{\dot{a}}{a}\right)^2+\frac{1}{a^2}
    &\cong \frac{1}{\alpha}\left[1-\left(1-\frac{2M\alpha}{a^{D-1}}-\alpha\Lambda\right)\left(1- \left(\frac{4\pi G_N\sqrt{\alpha}(\rho+\sigma)}{D-2}\right)^2\right)\right]\\
    &\cong \frac{1}{\alpha}\left[1-\left(1-\frac{2M\alpha}{a^{D-1}}-\alpha \Lambda-\alpha \left(\frac{4\pi G_N(\rho+\sigma)}{D-2}\right)^2\right)\right]\nonumber\\
    &=\frac{2M}{a^{D-1}}+\Lambda+\left(\frac{4\pi G_N(\rho+\sigma)}{D-2}\right)^2.
  \end{align}  
This is precisely the Friedmann equation in Einstein gravity Eq.~\eqref{braneworld-Einstein}.

Next, we focus on the behavior at a small-scale factor.  If the scale factor is sufficiently small so that 
\begin{equation}
    \frac{2M\alpha}{a^{D-1}}\gg 1\, ,
\end{equation}
then the right-hand side of Eq.~\eqref{eq-modified-Friedmann-Dymnikova-like} tends to zero (in fact, in this case, it vanishes exponentially fast).
Therefore, in this region, Eq.~\eqref{eq-modified-Friedmann-Dymnikova-like} becomes
  \begin{align}
    \label{eq-de-Sitter}
    \left(\frac{\dot{a}}{a}\right)^2+\frac{1}{a^2}-\frac{1}{\alpha}=0.
  \end{align}
  This is exactly the Friedmann equation with a positive cosmological constant $3/\alpha$ and can be solved as 
  \begin{align}
    \label{eq-inflation-de-Sitter}
    a=\sqrt{\alpha}\cosh{\left(\frac{\tau-\tau_{min}}{\sqrt{\alpha}}\right)}.
  \end{align}
Therefore, the brane behaves as de Sitter spacetime as long as $\frac{2M\alpha}{a^{D-1}}\gg 1$ holds.
This describes a bounce universe with a minimum scale factor $a_{\min}=\sqrt{\alpha}$. 
In order to gain more intuition about the different scales involved, it is useful to introduce the gravitational radius of the bulk black hole as\footnote{This is in general not the same as the horizon radius $r_{+}$, but it coincides with it when $|\Lambda|^{-1/2}\gg r_g\gg \sqrt{\alpha}$.}
\begin{equation}
r_g\coloneqq (2M)^{\frac{1}{D-3}}\, ,
\end{equation} 
which has dimensions of length. 
Taking into account that $a_{\min}=\sqrt{\alpha}$, we see there will be an inflationary phase as long as $r_g\gg \sqrt{\alpha}$, \textit{i.e.}, we only need to demand the bulk gravitational radius to be much larger than the length scale of new physics $\sqrt{\alpha}$, which is a very natural requirement. Since de Sitter inflation occurs as long as  $\frac{2M\alpha}{a^{D-1}}\gg1$, an order-of-magnitude estimate for the end of inflation is obtained by $\frac{2M\alpha}{a^{D-1}}\cong 1$, which yields
\begin{align}
    \label{eq-order-estimate}
    \frac{a_{\rm end}}{\sqrt{\alpha}}\cong \left(\frac{r_g}{\sqrt{\alpha}}\right)^{\frac{D-3}{D-1}}.
\end{align}
Since the minimum value of the scale factor is $a_{\min}=\sqrt{\alpha}$, this expression coincides with the definition of the number of e-folds $N\coloneqq\ln{\left(\frac{a_{\rm end}}{a_{\min}}\right)}$. Namely, in this model, the number of e-folds can be estimated as
\begin{align}
  \label{eq-e-folds}
    N\cong\frac{D-3}{D-1}\ln{\left(\frac{r_g}{\sqrt{\alpha}}\right)}.
\end{align}
Also, we can estimate the duration of the de Sitter phase by substituting Eq.~\eqref{eq-order-estimate} into Eq.~\eqref{eq-inflation-de-Sitter}, which yields $\tau_{f}\cong \sqrt{\alpha}N$, assuming that $N$ is large. 

An essential feature of these estimates is that these quantities $N$ and $\tau_f$ depend only on the bulk black hole mass $M$ and on the length scale of Birkhoff-QTG $\sqrt{\alpha}$ --- thus, they are completely independent of the matter content of the brane. In the standard cosmology, one typically requires $N\sim 60$ e-folds. To achieve this value, in this model, 
\begin{align}
  \frac{r_g}{\sqrt{\alpha}}\cong e^{120}
\end{align}
is needed. As we mentioned above, this condition can be naturally achieved by taking the bulk black hole to a large size. 

  \begin{figure}[t]
    \centering
    \includegraphics[width=0.48\linewidth]{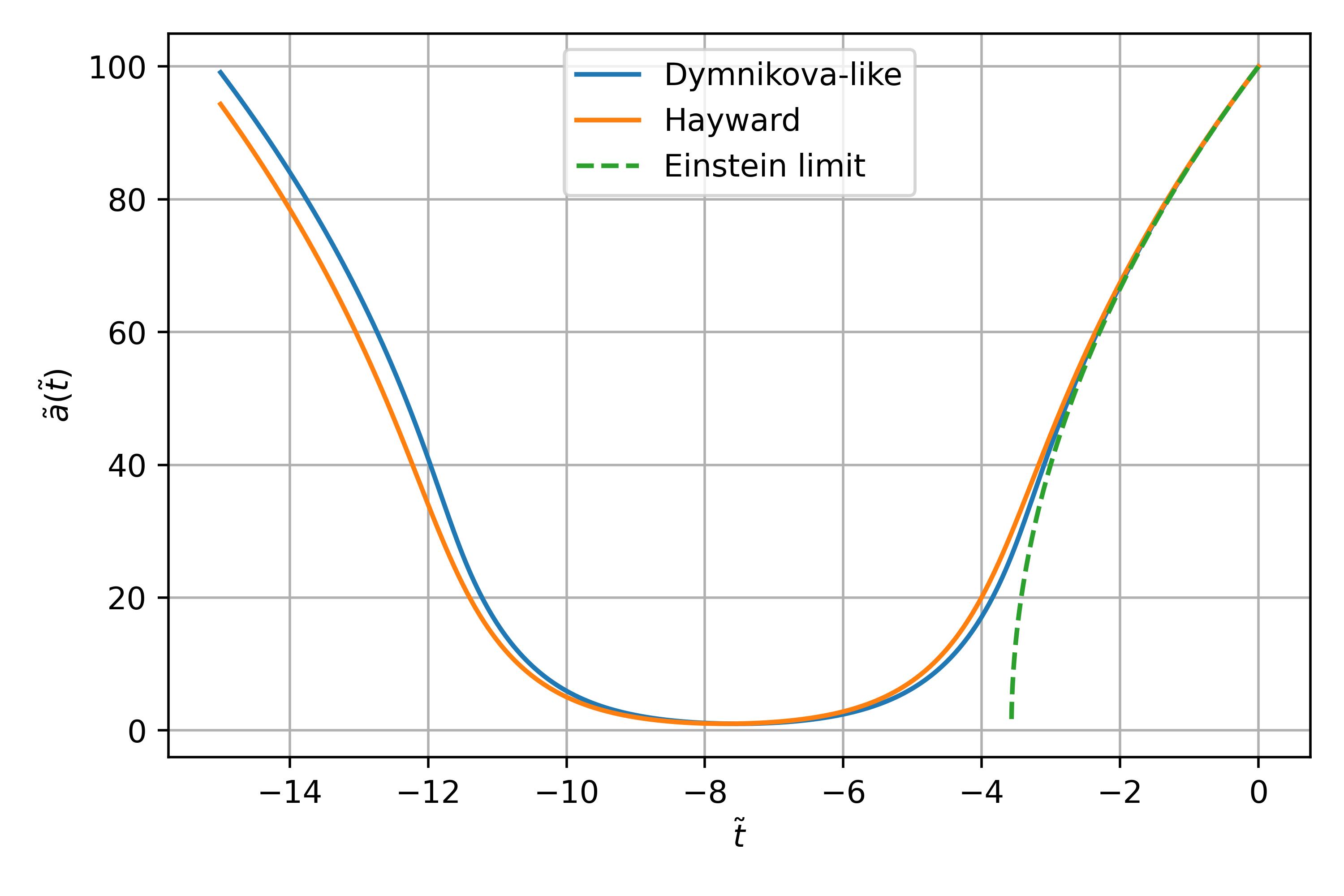}
    \includegraphics[width=0.48\linewidth]{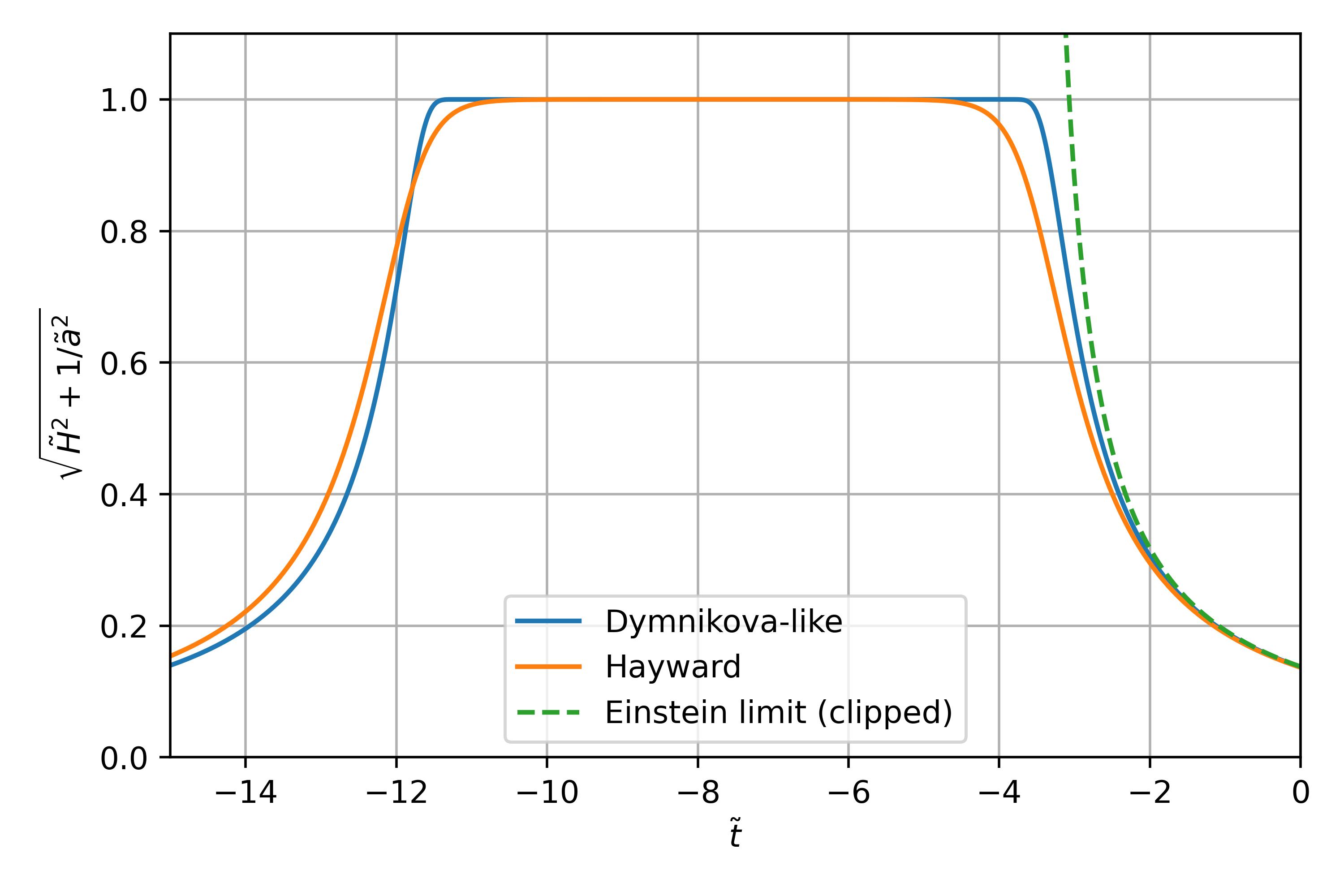}
    \caption{Scale factor $a$ and indicator $F=\sqrt{\tilde{H}^2+\frac{1}{\tilde{a}^2}}$ for parameters $D=5$, $\tilde{M}=10^9$, $\tilde{\sigma}=10^{-2}$, $\tilde{\Lambda}=-10^{-9}$, and $\tilde{\rho}=\frac{1}{\tilde{a}^4}$.  We start the evolution at a given time $\tilde{t}=0$ for which $\tilde a(0)=100$ and solve for the evolution of the scale factor backwards in time. The blue curves represent the Dymnikova-like braneworld scenario, the orange lines represent the Hayward braneworld scenario, and the green dashed curves represent an Einstein-gravity limit. While all the curves coincide at late times, they disagree for earlier times: Einstein gravity predicts a singularity in the past, and the other two models a de Sitter phase with a bounce. }
    \label{pic-numerical}
  \end{figure}

\subsection{Hayward scenario}
Let us now choose the bulk gravitational theory as the one giving rise to the regular Hayward black hole in Eq.~\eqref{eq-Hayward-f}. In this case, the modified Friedmann equation becomes
  \begin{align}
    \label{eq-modified-Friedmann-Hayward}
    \left(\frac{\dot{a}}{a}\right)^2+\frac{1}{a^2}-\frac{1}{\alpha}=-\frac{\cos^2{\Theta}}{\alpha\left[1+\alpha\Lambda+\frac{2M\alpha}{a^{D-1}} \right]},
  \end{align}
where
  \begin{align}
    \begin{cases}
      \Theta\coloneqq \frac{8\pi G_N\alpha^2}{D-2}(\rho+\sigma)A^3-\sqrt{\alpha} A\left(1+\alpha\Lambda+\frac{2M\alpha}{a^{D-1}}\right)\sqrt{\frac{1}{1+\alpha\Lambda+\frac{2M\alpha}{a^{D-1}}}-\alpha A^2},\\
    A\coloneqq \sqrt{\frac{1}{\alpha}-\left(\frac{\dot{a}}{a}\right)^2-\frac{1}{a^2}}.
    \end{cases}
  \end{align}
  The derivation is also presented in the Appendix~\ref{Appendix-proof}. We refer to this as a Hayward braneworld scenario. It is easy to check that in the low-energy regime, Eq.~\eqref{eq-modified-Friedmann-Hayward} reproduces the  Einstein gravity Friedmann equation Eq.~\eqref{eq-braneworld-cosmology}. Moreover, for $\frac{2M\alpha}{a^{D-1}}\gg1$, we have a very similar behavior to the Dymnikova-like scenario: the right-hand side of Eq.~\eqref{eq-modified-Friedmann-Hayward} vanishes and the brane approaches the closed de Sitter solution. In particular, the number of e-folds can again be estimated  by Eq.~\eqref{eq-e-folds}. 

To see the full evolution of the scale factor and to validate the estimation of the number of e-folds, we solve the modified Friedmann equations Eq.~\eqref{eq-modified-Friedmann-Dymnikova-like}  and Eq.~\eqref{eq-modified-Friedmann-Hayward} numerically. To do this, we introduce dimensionless quantities, denoted by a tilde, namely
\begin{align}
  \tilde{a}\coloneqq \frac{a}{\sqrt{\alpha}},\quad \tilde{t}\coloneqq \frac{\tau}{\sqrt{\alpha}},\quad \tilde{M}\coloneqq \frac{M}{\alpha^{\frac{D-3}{2}}},\quad \tilde{\Lambda}\coloneqq \alpha\Lambda,\quad \tilde{\rho}\coloneqq \frac{4\pi G_N \sqrt{\alpha}}{D-2}\rho,\quad \tilde{\sigma}\coloneqq \frac{4\pi G_N\sqrt{\alpha}}{D-2}\sigma.
\end{align}
In this setup, in Figure \ref{pic-numerical}, we plot the scale factor $\tilde{a}$, and an indicator $F=\sqrt{\tilde{H}^2+\frac{1}{\tilde{a}^2}}$ that takes the value $F=1$ for exact de Sitter phase where $\tilde{H}\coloneqq \frac{\dot{\tilde{a}}}{\tilde{a}}$ is the dimensionless Hubble parameter. 
Figure \ref{pic-numerical} shows that the Hayward and Dymnikova-like scenarios clearly develop a quasi-de Sitter phase corresponding to the region where $F$ is almost flat. The number of e-folds, estimated by identifying the time at which $\ddot{a}=0$, is approximately $N_{\rm Dymnikova}=3.57$ and $N_{\rm Hayward}=3.63$, for the parameters used to produce Figure \ref{pic-numerical}.  These results agree almost perfectly with the theoretical order-of-magnitude estimate Eq.~\eqref{eq-e-folds}, which yields $N=3.63$ for $\tilde M=10^9$ and $D=5$.  

\subsection{Universality of de Sitter inflation and the number of e-folds}
In the two explicit examples above, we observed that in the small scale factor regime where $\frac{2M\alpha}{a^{D-1}}\gg1$, the brane evolution approaches the closed de Sitter behavior Eq.~\eqref{eq-de-Sitter} essentially independently of the bulk black hole mass and the brane matter content. In this section, we show that this early-time de Sitter behavior is universal: it holds for any infinite-tower Birkhoff-QTG satisfying Eq.~\eqref{condition-to-parameter} once $\mathbb{Z}_2$ symmetry is imposed. We also derive a universal order-of-magnitude estimation for the number of e-folds.

First, we prove the universality of de Sitter inflation. To this end, we consider the non-negative quantity\footnote{We recall that, in fact, $a_n>0$ follows from the first Friedmann equation Eq.~\eqref{eq-braneworld-EOM-1} upon the assumption that $\rho+\sigma>0$.} $a_n^2=f+\dot{a}^2$, which can be written as 
\begin{align}
  0\le \frac{a_n^2}{a^2}&=\left[\frac{1}{\alpha}-\frac{1-f(a)}{a^2}\right]-\left[\frac{1}{\alpha}-\frac{1}{a^2}-\left(\frac{\dot{a}}{a}\right)^2\right]\, .
  \label{eq-condition-timelike}
\end{align}
Then, we can see that the two terms between brackets in the right-hand side are positive from the inequalities Eq.~\eqref{eq-bulk-constraint} and Eq.~\eqref{eq-braneworld-constraint}. Due to the universal behavior of the function $f$ that we saw in Eq.~\eqref{eq-small-r-f}, for $a\rightarrow0$, the first bracket in Eq.~\eqref{eq-condition-timelike} tends to zero.
If the second bracket tended to a positive constant for $a\to 0$, then for sufficiently small $a$ we would have $\frac{a_n^2}{a^2}<0$, which contradicts $a_n^2\geq 0$.
Hence, the second bracket must tend to zero, which gives Eq.~\eqref{eq-de-Sitter}, de Sitter inflation. This statement is based on the inequality Eq.~\eqref{eq-braneworld-constraint}, which originates from imposing $\mathbb{Z}_2$ symmetry. In other words, $\mathbb{Z}_2$ symmetry in the context of regular black holes in Birkhoff-QTG generally leads to de Sitter inflation. 

Next, we prove the universal expression of the number of e-folds. To prove it, let us introduce a new function $g(x)$ defined by
\begin{equation}
g(x)=\alpha h(x/\alpha)\, .
\end{equation}
This is a dimensionless function of the dimensionless variable $x=\alpha\psi$, and since $h(\psi)$ diverges at $\psi=1/\alpha$, it follows that $g(x)$ diverges at $x=1$. 
We can then write the master equation Eq.~\eqref{eq-h-Regular-black-hole} in a dimensionless way as
\begin{align}
  \label{eq-EOM}
  g(x)=\frac{2M\alpha}{r^{D-1}}\, .
\end{align}
Inverting this equation, and taking into account that $x=\alpha(1-f)/r^2$, we obtain 
\begin{align}
  \label{eq-f-with-g}
  f(r)&=1-g^{-1}\left(\frac{2M\alpha}{r^{D-1}}\right)\frac{r^2}{\alpha}.
\end{align}
Since $g(x)$ has radius of convergence 1, the inverse function satisfies $g^{-1}(z)\to 1$ when $z= \frac{2M\alpha}{r^{D-1}}\to \infty$. This means that the metric function $f(r)$ acquires the de Sitter behavior Eq.~\eqref{eq-small-r-f} whenever $\frac{2M\alpha}{r^{D-1}}\gg 1$. 
As a consequence of the inequality Eq.~\eqref{eq-condition-timelike}, this also implies that the evolution of the brane is governed by the de Sitter phase Eq.~\eqref{eq-de-Sitter} whenever $\frac{2M\alpha}{r^{D-1}}\gg 1$, and that inflation ends when
\begin{align}
  \frac{2M\alpha}{a^{D-1}}\sim 1\, ,
\end{align}
up to an order-one constant. This is equivalent to the condition we obtained in the Hayward and Dymnikova-like scenarios, and it implies that the number of e-folds is universally given by 
\begin{align}
  \label{eq-e-folds-general}
  N\cong \frac{D-3}{D-1}\ln{\left(\frac{r_g}{\sqrt{\alpha}}\right)}\, ,
\end{align}
independently of the specific choice of the coupling constants $\{\alpha_n\}$.

\section{Conclusion and discussion}
\label{sec-Conclusion}

In this work, we have investigated braneworld cosmology in Birkhoff-QTG with an infinite tower of higher-curvature corrections, placing particular emphasis on scenarios in which the higher-dimensional bulk admits regular black hole solutions. A particularly interesting feature of this construction is that it allows us to connect QT gravities --- which are only defined in $D\ge 5$ --- to four-dimensional physics. 

Starting from the $\mathbb{Z}_2$-symmetric junction conditions for a brane moving on a static and spherically symmetric bulk, we derived the modified Friedmann equations governing a closed FLRW braneworld. We verified that the standard Einstein-gravity braneworld equations are recovered in the appropriate low-energy regime, providing a consistency check and clarifying how the higher-curvature corrections become relevant at small scale factors.  

Our first result is that, within the scenarios with regular black holes, the braneworld is generally driven toward a closed de Sitter phase governed by Eq.~\eqref{eq-de-Sitter}. This happens when the brane is inside the de Sitter core of the bulk black hole, corresponding to the regime $\frac{2M\alpha}{a^{D-1}}\gg1$. In addition, rather than a big bang singularity, the universe experiences a bounce when the scale factor reaches the scale of new physics.\footnote{A softening of the big bang singularity was previously observed in braneworld models with quadratic curvature corrections \cite{Kofinas:2003rz}, but in our case the singularity is completely removed thanks to the infinite tower of corrections.} The approach to de Sitter is independent of the matter content of the brane and of the detailed higher-curvature couplings $\{\alpha_n\}$, as long as they satisfy the general requirements in Eq.~\eqref{condition-to-parameter}. This early de Sitter behavior is a genuine bulk-induced effect that crucially relies on the presence of the regular bulk black hole. 

Our second result is a universal estimate for the number of e-folds of inflation that only depends on the ratio of the bulk black hole scale $r_g$ and the length parameter of Birkhoff-QTG $\sqrt{\alpha}$  --- see Eq.~\eqref{eq-e-folds-general}. Numerical solutions of the modified Friedmann equations for two explicit bulk regular black holes (Dymnikova-like and Hayward black holes) confirm the analytic estimates and demonstrate that the de Sitter phase and the bounce emerge robustly across different choices of the all-order couplings and bulk geometries. 

As we mentioned in the introduction, inflationary solutions can also arise in infinite-order GQTGs in four-dimensions, giving rise to the geometric inflation scenario \cite{Arciniega:2018fxj,Arciniega:2018tnn,Cisterna:2018tgx}.  However, an obstacle in the original geometric inflation proposal is that achieving a sufficient number of e-folds typically requires trans-Planckian matter energy densities, raising concerns about the regime of validity of the effective description \cite{Edelstein:2020nhg}. A crucial advantage of the braneworld setup presented here is that the number of e-folds depends only on the ratio of bulk scales $r_g/\sqrt{\alpha}$ and not on the matter content of the brane. In other words, inflation happens universally and independently of the matter content. This means that, by choosing an appropriate theory for the matter description, one could start with an ``empty'' inflationary universe --- hence avoiding trans-Planckian energy densities --- and later generate a matter density via an appropriate reheating mechanism. As inflation takes place literally for \emph{any} matter content, there are potentially many ways in which this can be achieved.   

Another noticeable feature of the solutions presented here is that the inflationary phase shows a constant Hubble radius --- this is evident from the right panel in Fig.~\ref{pic-numerical}. Therefore these models could be labeled as ultra-slow-roll inflation, or even, no-roll inflation. Interestingly, it is also possible to realize slow-roll scenarios --- in which the Hubble radius runs as a function of the number of e-folds --- within the setup of Birkhoff-QTG braneworld cosmology. To this end, we simply need to relax the conditions on the QTG couplings \eqref{condition-to-parameter} to allow the function $h(\psi)$ to have an infinite radius of convergence (rather than a finite one). In these situations, one can obtain bulk solutions that behave as, \textit{e.g.}, $f(r)\approx 1-r^2/\alpha \log(r/\sqrt{\alpha})$ for $r\to 0$ \cite{Bueno:2024dgm}. While such geometries are not fully regular at $r=0$, they can be interpreted as a quasi-de Sitter spacetime with a running Hubble parameter. We have checked that, in these cases, the braneworld theory does indeed show a slow-roll inflationary phase. We leave the analysis of these models for future work.    

There are several directions that merit further investigation. First, it is essential to study cosmological perturbations around the background solutions obtained here in order to assess whether the scenario can reproduce the observed spectrum of primordial fluctuations and to identify distinctive signatures of bulk-induced inflation. Second, our analysis assumes $\mathbb{Z}_2$ symmetry and a static bulk mass parameter; relaxing these assumptions, allowing for asymmetric embeddings \cite{Davis:2000jq,Gergely:2003pn,Padilla:2004tp,Konya:2006wr,Yamauchi:2007wm}, brane-bulk energy exchange, or time-dependent bulk black hole mass would clarify how general the inflationary mechanism remains beyond the present setup.

\section*{Acknowledgements}
We would like to thank Pablo Bueno, Robie Hennigar and \'Angel Murcia for useful comments and discussions. 
K.S. is in part supported by JST, the establishment of university fellowships towards the creation of science technology innovation, Grant Number JPMJFS2138.
The work of P.A.C. is supported by a Ram\'on y
Cajal fellowship (RYC2023-044375-I) and by a Proyecto de generaci\'on de conocimiento (PID2024-155685NB-C22) from Spain’s
Ministry of Science, Innovation and Universities.

 \appendix
 \def\thesection{\Alph{section}} 
 \section{Derivation of modified Friedmann equation for concrete models}
 \label{Appendix-proof}
\subsection{Dymnikova-like scenario}
  The Dymnikova-like black hole can be obtained when $\alpha_n=\frac{\alpha^{n-1}}{n}$
  is satisfied. In this situation, $h(\psi)$ and $h'(\psi)$ satisfy $h(\psi)=-\Lambda-\frac{\ln{(1-\alpha \psi)}}{\alpha}, h'(\psi)=\frac{1}{1-\alpha \psi}$, the integral of the junction condition can be evaluated as
  \begin{align}
    L&\coloneqq \int_{0}^{a_n}{dz h'\left(\frac{1+\dot{a}^2-z^2}{a^2}\right)}\nonumber\\
    &=\frac{1}{\alpha}\int_0^{a_n}\frac{dz}{\frac{z^2}{a^2}+\left[\frac{1}{\alpha}-\left(\frac{\dot{a}}{a}\right)^2-\frac{1}{a^2}\right]}\nonumber\\
    &=\frac{a}{\alpha\sqrt{\frac{1}{\alpha}-\left(\frac{\dot{a}}{a}\right)^2-\frac{1}{a^2}}}\arctan{\left(\frac{\sqrt{\left(\frac{\dot{a}}{a}\right)^2+\frac{f}{a^2}}}{\sqrt{\frac{1}{\alpha}-\left(\frac{\dot{a}}{a}\right)^2-\frac{1}{a^2}}}\right)},
  \end{align}
    where we used a formula
  \begin{align}
    \int{\frac{dx}{\frac{x^2}{a^2}+A^2}}=\frac{a}{A}\arctan{\left(\frac{x}{aA}\right)}.
  \end{align}
  From this expression, the junction condition Eq.~\eqref{eq-braneworld-EOM-1} is expressed as 
  \begin{align}
    \frac{\sqrt{\left(\frac{\dot{a}}{a}\right)^2+\frac{f}{a^2}}}{\sqrt{\frac{1}{\alpha}-\left(\frac{\dot{a}}{a}\right)^2-\frac{1}{a^2}}}=\tan{\left(\sqrt{\frac{1}{\alpha}-\left(\frac{\dot{a}}{a}\right)^2-\frac{1}{a^2}}\frac{4\pi G_N\alpha(\rho+\sigma)}{D-2}\right)}.\label{eq-Dymnikova-pre}
  \end{align}
  Using Eq.~\eqref{eq-braneworld-constraint}and $f+\dot{a}^2>0$, we can rewrite Eq.~\eqref{eq-Dymnikova-pre} as
\begin{align}
      \left(\frac{\dot{a}}{a}\right)^2+\frac{1}{a^2}-\frac{1}{\alpha}=-\frac{1}{\alpha}\exp{\left[-\frac{2M\alpha}{a^{D-1}}-\alpha \Lambda\right]}\cos^2{\left[\sqrt{-\left(\frac{\dot{a}}{a}\right)^2-\frac{1}{a^2}+\frac{1}{\alpha}}\cdot \frac{4\pi G_N(\rho+\sigma)\alpha}{D-2}\right]}.\nonumber
  \end{align}

\subsection{Hayward scenario}
  When the coupling constants satisfy $\alpha_n=\alpha^{n-1}$, $h(\psi)$ and $h'(\psi)$ satisfy $h(\psi)=-\Lambda+\frac{\psi}{1-\alpha\psi}, h'(\psi)=\frac{1}{(1-\alpha\psi)^2}$, the integral of the junction condition can be evaluated as
  \begin{align}
    L&\coloneqq \int_{0}^{a_n}{dz h'\left(\frac{1+\dot{a}^2-z^2}{a^2}\right)}\nonumber\\
    &=\frac{1}{\alpha^2}\int_0^{a_n}{\frac{dz}{\left[\frac{z^2}{a^2}+\left[\frac{1}{\alpha}-\left(\frac{\dot{a}}{a}\right)^2-\frac{1}{a^2}\right]\right]^2}},\nonumber
  \end{align}
  where, since the following formula 
  \begin{align}
    \int{\frac{dz}{\left(\frac{z^2}{a^2}+A^2\right)^2}}=\frac{a}{2A^3}\left[\frac{aAz}{a^2A^2+z^2}+\arctan{\left[\frac{z}{aA}\right]}\right]
  \end{align}
  holds, $L$ can be estimated with $A\coloneqq \sqrt{\frac{1}{\alpha}-\left(\frac{\dot{a}}{a}\right)^2-\frac{1}{a^2}}$ as
  \begin{align}
      L&=\frac{a}{2\alpha^2A^3}\left[\frac{aAz}{a^2A^2+z^2}+\arctan{\left[\frac{z}{aA}\right]}\right]_{0}^{a_n}\nonumber\\
      &=\frac{a}{2\alpha^2A^3}\left[\frac{aA\varphi_n}{a^2A^2+\varphi_n^2}+\arctan{\left[\frac{\varphi_n}{aA}\right]}\right].
  \end{align}
From this expression, the junction condition Eq.~\eqref{eq-braneworld-EOM-1} is expressed as 
  \begin{align}
    \label{eq-Hayward-pre}
    \frac{\sqrt{\left(\frac{\dot{a}}{a}\right)^2+\frac{f}{a^2}}}{\sqrt{\frac{1}{\alpha}-\left(\frac{\dot{a}}{a}\right)^2-\frac{1}{a^2}}}&=\tan{\left[\frac{8\pi G_N\alpha^2A^3(\rho+\sigma)}{D-2}-\frac{A\sqrt{\left(\frac{\dot{a}}{a}\right)^2+\frac{f}{a^2}}}{\frac{1}{\alpha}-\frac{1-f}{a^2}}\right]}.
  \end{align}
Since 
\begin{align}
  \frac{1}{\alpha}-\frac{1-f}{a^2}=\frac{1}{\alpha}\left[\frac{1}{1+\alpha\Lambda+\frac{2M\alpha}{a^{D-1}}}\right]
\end{align}
and 
\begin{align}
  \left(\frac{\dot{a}}{a}\right)^2+\frac{f}{a^2}&=\left(\frac{\dot{a}}{a}\right)^2+\frac{1}{a^2}-\frac{1}{\alpha}+\frac{1}{\alpha}\left[\frac{1}{1+\alpha\Lambda+\frac{2M\alpha}{a^{D-1}}}\right]
\end{align}
 hold, we obtain
 \begin{align}
   \Theta&\coloneqq \frac{8\pi G_N\alpha^2A^3(\rho+\sigma)}{D-2}-\frac{A\sqrt{\left(\frac{\dot{a}}{a}\right)^2+\frac{f}{a^2}}}{\frac{1}{\alpha}-\frac{1-f}{a^2}}\\
   &=\frac{8\pi G_N\alpha^2}{D-2}(\rho+\sigma)A^3-\sqrt{\alpha} A\left(1+\alpha\Lambda+\frac{2M\alpha}{a^{D-1}}\right)\sqrt{\frac{1}{1+\alpha\Lambda+\frac{2M\alpha}{a^{D-1}}}-\alpha A^2}.
 \end{align}
  Using Eq.~\eqref{eq-braneworld-constraint} and $\Theta$, we can rewrite Eq.~\eqref{eq-Hayward-pre} as
  \begin{align}
        \left(\frac{\dot{a}}{a}\right)^2+\frac{1}{a^2}-\frac{1}{\alpha}=-\frac{\cos^2{\Theta}}{\alpha\left[1+\alpha\Lambda+\frac{2M\alpha}{a^{D-1}} \right]}.  
  \end{align}

	\bibliographystyle{JHEP}
	\bibliography{Regular-v2.bib}

\end{document}